\documentclass[draft,onecolumn]{IEEEtran}
\usepackage{Cocco2026_Properties_CQ_Ensembles}
\usepackage{flushend}
\usepackage[utf8]{inputenc}

\usepackage{braket}

\usepackage{url}

\usepackage[ 
top    = 0.75in,
bottom = 1.1in,
left   = 0.63in,
right  = 0.63in]{geometry}

\title{Properties of Random Code Ensembles over Classical-Quantum Channels}

\author{Giuseppe Cocco and Javier Rodríguez Fonollosa
\thanks{Giuseppe Cocco and Javier Rodríguez Fonollosa are with the Department of Signal Theory and Communications, Universitat Polit\`ecnica de Catalunya, 08034, Barcelona, Spain (Corresponding author: giuseppe.cocco@upc.edu).

  This work was supported in part by the Ramon y Cajal fellowship program, grant RYC2021-033908-I, and grant PID2022-137099NB-C41 both funded by 
MCIN/AEI/10.13039/501100011033/FEDER, the European Union ``NextGenerationEU'' Recovery Plan for Europe, and by grant 2021 SGR 01033 funded by AGAUR, Departament de Recerca i Universitats de la Generalitat de Catalunya 10.13039/501100002809.\\

Part of the content of this work has been presented at the IEEE Information Theory Workshop 2025 (IEEE ITW 2025), Sep. 29 - Oct., Sydney, Australia.
}
	}

\begin{document}
\maketitle

\begin{abstract}
We show two properties of i.i.d. code ensembles over classical-quantum (CQ) channels with arbitrary output states.

The first property is that the probability distribution of the error exponent across the ensemble accumulates above a threshold that is strictly larger than the CQ random coding exponent (RCE) at low rates, while coinciding with it at rates close to the mutual information of the channel. 
This result, combined with the works by Dalai \cite{sp_cq_dalai_2023}, Renes \cite{renesTIT2025_tight_LB}, Li and Yang \cite{reliability_CQ_Li_Yang_2025} and Cheng and Liu \cite{cheng2025errorexponentsquantumpacking}, implies that the ensemble  distribution of error exponents concentrates around the CQ RCE in the high-rate regime. Moreover, in the same rate region the threshold we derive coincides with the ensemble-average of the exponent, that is, the CQ typical random coding (TRC) exponent.

The second property establishes that the probability of a randomly selected code containing a nested code with the same rate, such that each of its codewords achieves the CQ expurgated exponent, goes to one asymptotically in the codeword length. The nested code can be obtained by expurgating a fraction of codewords which is vanishingly small as the codeword length grows. This result refines Holevo's work \cite{Q_expur_general_Holevo200}, which proved that the expurgated exponent can be achieved by discarding the worst half of the codewords from at least one code in the ensemble.
\end{abstract}
\vspace{-0.1in}	

\section{Introduction}\label{sec:intro}

\subsection{Preliminaries}\label{sec:prelim}
Let $\mathcal{H}_A$ be a Hilbert space of dimension $d_A$ and $\tau_A\in \mathcal{D}(\mathcal{H}_A)$ the density matrix of a quantum state, $\mathcal{D}(\mathcal{H}_A)$ being the set of all density operators acting on $\mathcal{H}_A$. Let $\mathcal{H}_B$, $d_B$, $\sigma_B$ and $\mathcal{D}(\mathcal{H}_B)$ be similarly defined. 
In the most general form of \ac{CQ} channels \cite{wildeBook_2021}, an input state $\tau_A$ is measured using an orthonormal basis, and a density operator conditioned on the result of such measurement is found at the channel output. Specifically, given an input state $\tau_A$, an orthonormal basis $\{|x\rangle_A\}_{x\in\mathcal{X}}$ in $\mathcal{H}_A$, where $\mathcal{X}$ is an index set with $|\mathcal{X}| = d_A$, and a set of generic (not necessarily pure) states $\{\sigma^x_B\}_{x\in\mathcal{X}}$ belonging to $\mathcal{D}(\mathcal{H}_B)$, a \ac{CQ} channel operates on $\tau_A$ as follows
\begin{align}
\mathcal N_{A \rightarrow B}(\tau_A)=\sum_{x\in\mathcal{X}} \langle x|_A\tau_A|x\rangle_A \sigma^x_B,
\end{align}
where $\langle x|_A\tau_A|x\rangle_A$ represents the probability of measuring the outcome with index $x$ and hence $\sum_{x\in\mathcal{X}}\langle x|_A\tau_A|x\rangle_A=1$. This general model can be specialized as follows: given a set $\mathcal{X}=\{1,\ldots,|\mathcal{X}|\}$ of classical symbols, in order to transmit symbol $x\in\mathcal{X}$ the pure state $\ket{x}_A$ is selected from the orthonormal set of state vectors and used as input to the quantum channel. This yields an output state with density matrix $\sigma^x_B$. 
Consequently, the CQ channel maps a classical input symbol $x$ to a quantum output state $\sigma_B^x$.

Accordingly, a sequence of symbols $\x=(x_1,x_2,\ldots,x_n)$ has an associated input state $\ket{\x}=\ket{x_1}\otimes\ket{x_2}\otimes\ldots \otimes\ket{x_n}\in \mathcal{H}_A^{\otimes n}$, which induces the output state $\sigma_B^{\x}=\sigma_B^{x_1}\otimes\sigma_B^{x_2}\otimes\ldots \otimes\sigma_B^{x_n}\in \mathcal{D}(\mathcal{H}_B^{\otimes n})$. Given the deterministic mapping between $\x$ and $\sigma_B^{\x}$, the \ac{CQ} channel can be equivalently considered to have a classical input $\x$ and a quantum output $\sigma_B^{\x}$. In the sequel, subscript $B$ will be omitted unless needed.

A channel code $\CnallMdet$ for such channel is a mapping from a set of $M_n=2^{nR}$ messages $\mathcal{M}_n=\{1,\ldots,M_n\}$ to a set of codewords $\{\x_1,\x_2,\ldots,\x_{M_n}\}$, $\x_j\in\mathcal{X}^n,  \forall j\in\{1,\ldots,M_n\}$, $R=\frac{1}{n}\log M_n$ being the rate of the code\footnote{In this definition we use a slight abuse of notation since $M_n$ is an integer and thus $R$ is a function of $n$. We omit such dependence in the following to limit visual clutter.}. All logarithms are assumed to be in base $2$.

A quantum decision rule for such code is a \ac{POVM} \cite{math_foun_QMech_VonNeumann2018} $\{\Pi_0, \Pi_1,\Pi_2,\ldots,\Pi_{M_n}\}$, where $\Pi_i$, $i=0, \ldots, M_n$, are positive semi-definite linear operators in $\mathcal{H}^{\otimes n}$ that satisfy completeness, that is $\Pi_i\succeq 0$ $\forall i$ and $\sum_{m=0}^{M_n} \Pi_m= I$, where $I$ is the identity matrix and $\Pi_0=I-\sum_{m=1}^{M_n} \Pi_m$. 
The \ac{POVM} is used to measure the output of the \ac{CQ} channel. When codeword $\x_m$ is transmitted, the probability that, for the chosen \ac{POVM}, the measurement yields the result $\hat{m}$ is given by $\text{Pr}\{\hat{m}|m\}=\tr\{\Pi_{\hat{m}}\sigma^{\x_m} \}$. Thus, for a given code $\CnallMdet$ the probability of error when transmitting $\x_m, m\in\{1,\ldots,M_n\}$, is:
\begin{align}\label{eq:pe_m}
\PecnmallMdet = 1- \tr\{\Pi_{m}\sigma^{\x_m} \}.
\end{align}
Using \eqref{eq:pe_m}, and assuming equally likely codewords, the average probability of error for code $\CnallMdet$ is:
\begin{align}\label{eq:pe_code}
\PecnallMdet &= \frac{1}{M_n}\sum_{m=1}^{M_n}\PecnmallMdet\\
&= 1 - \frac{1}{M_n}\sum_{m=1}^{M_n}\tr\{\Pi_{m}\sigma^{\x_m} \}.
\end{align}
Let $\CnallM=\{\X_1, \dotsc, \X_{M_n}\}$ denote a random code from an \ac{i.i.d.}  ensemble, so that codewords are generated independently according to a distribution $Q^n(\X)=\prod_{i=1}^n Q(X_i)$, $Q(X)$ being the symbol distribution defined over $\mathcal{X}$. To simplify the notation, in the following we will use $\Cn$ in place of $\CnallM$.

Let $\EE[\Pecn]$ be the ensemble average of the error probability and let $-\frac{1}{n}\log\EE[\Pecn]$ be its negative and normalized exponent. 
For classical channels, Gallager derived an upper bound on $\EE[\Pecn]$~\cite[Eq.~(5.6.16)]{gallagerBook} and called its exponent the \ac{RCE}. An analogous upper bound on the ensemble-average error  probability for random i.i.d. codes in \ac{CQ} channels was derived for the case of pure quantum states by Burnashev and Holevo  \cite{Q_randomCoding_Brunashev1998}. In  \cite{Q_randomCoding_Brunashev1998} the authors also conjectured that the same expression  holds for \ac{CQ} channels with arbitrary (mixed) states, showing that this is indeed the case for some specific setup. 
In \cite{sp_cq_dalai_2023} Dalai derived the sphere-packing upper bound on the largest exponent in the ensemble (the reliability function \cite[Chapter 5]{gallagerBook}) for generic \ac{CQ} channels and pointed out that it coincides with Holevo's pure-state \ac{RCE} at rates close to capacity.
Renes \cite{renesTIT2025_tight_LB} and Li and Yang \cite{reliability_CQ_Li_Yang_2025} independently proved that Holevo's conjectured expression for the \ac{RCE} of generic channels is at the same time a lower bound and an upper bound on the reliability function.
These results imply that at high rates the \ac{RCE} is in fact the reliability function for general \ac{CQ} channels.  
Cheng and Liu \cite{cheng2025errorexponentsquantumpacking} recently proved that Holevo's expression matches the exact exponent of the ensemble-average error probability for generic \ac{CQ} channels.

The \ac{RCE} for \ac{CQ} channels with general quantum states is \cite{renesTIT2025_tight_LB}\cite{Q_expur_general_Holevo200}:
\begin{align}\label{eq:QEr}
\Er(R,Q) = \max_{0\leq s\leq 1} -\log \left(\Tr \left[\left(\sum_{x\in\mathcal{X}}Q(x)\left(\sigma^x\right)^{\frac{1}{1+s}}\right)^{1+s}\right]
\right) - sR.
\end{align}
The exponent $\Er(R,Q)$ is strictly positive at rates below the mutual information of the channel for a given input distribution $Q(x)$. If the input distribution is optimized, the exponent is positive for rates below the classical capacity of the \ac{CQ} channel $C$ given by \cite[Theorem 1]{holevo_capacity_quantum_1998}\cite{Schumacher_quantum_capacity_1997}:
\begin{align}\label{eq:capacity}
    C=\max_{Q(x)} \left[H\left(\sigma\right)-\sum_{x\in\mathcal{X}}Q(x)H(\sigma^x)\right]
\end{align}
where $\sigmaav=\sum_{x\in\mathcal{X}}Q(x)\sigma^x$ while $H(\sigmaav)$ is the Von Neumann entropy of $\sigmaav$, defined as:
\begin{align}
H(\sigma) &= -\text{Tr}\{\sigmaav\log \sigmaav\}\\
&=-\sum_{i=1}^d\lambda_i\log \lambda_i
\end{align}
and $\{\lambda_1,\ldots,\lambda_d\}$ are the eigenvalues of $\sigmaav$. Notice that, while for general quantum channels a regularization is needed in the definition of capacity, this is not the case in \eqref{eq:capacity}. This is because the considered channel is a special case of an entanglement-breaking channel, for which the additivity of Holevo information holds and regularization is not needed \cite{wildeBook_2021}.

For classical channels  Gallager \cite{gallager1965simple}  showed that there exists at least one code with an error exponent larger than or equal to the so-called expurgated exponent. Such code can be obtained by removing half the codewords from a mother code having twice the number of desired codewords. For some channels and ensembles, at low rates the expurgated exponent is strictly larger than the \ac{RCE}. An analogous result is derived in \cite{Q_randomCoding_Brunashev1998} for \ac{CQ} channels with pure states, then generalized in \cite{Q_expur_general_Holevo200} to the case of generic  states.
The \ac{CQ} expurgated exponent is:
\begin{align}\label{eqn:expu}
	\Eex = \max_{\Grho\geq 1} \left\{E_{\rm x}(\Grho,Q) - \Grho R\right\},
\end{align}
where
\begin{align}\label{eqn:ex}
\Exvarrho =-\Grho\log \left( \sum_{x\in\mathcal{X}}\sum_{x'\in\mathcal{X}} Q(x)Q(x') \left( {\rm Tr}\left\{\sqrt{\sigma^x}\sqrt{\sigma^{x'}}\right\}\right)^\frac{1}{\Grho}  \right).
\end{align}

Recent years have witnessed a rising research interest in the \ac{TRC} exponent \cite{barg_forney_TIT2002}, defined as \cite{merhav_TIT2018}
\begin{align}\label{eq:trc_def}
\Etrccc\triangleq\lim_{n\rightarrow\infty}\EE\left[-\frac{1}{n}\log \Pecn\right].
\end{align}
The \ac{TRC} exponent captures the average exponent behavior of codes in random ensembles more accurately than the \ac{RCE}. Notice that the \ac{TRC} is larger than or equal to the \ac{RCE} since, by Jensen's inequality, it holds that:
\begin{align}
-\frac{1}{n}\log\EE[\Pecn] \leq \EE\left[-\frac{1}{n}\log \Pecn\right].
\end{align}
While the \ac{TRC} exponent has been studied in the context of classical communications, up to our knowledge it has not been investigated so far for \ac{CQ} channels.

\subsection{Contribution}
While previous results provide foundational knowledge on quantum code ensembles in \ac{CQ} channels with arbitrary output states, the distribution of the error exponent as such remains largely unstudied. In this work we derive two key properties of i.i.d. code ensembles over these channels.

\subsubsection{First Contribution}
We show that the mass of the exponent distribution for \ac{i.i.d.} code ensembles in \ac{CQ} channels lies above the quantity $\Etrccclb$, a lower bound on the \ac{TRC} exponent which we derive in the next section. In other words, the probability of finding a code with an exponent lower than $\Etrccclb$ is vanishingly small in the codeword length. At rates close to the mutual information of the  channel\footnote{Although we retain the dependence on the input distribution for the sake of generality, optimizing over it yields positive exponents up to the channel capacity.} $\Etrccclb$ coincides with the \ac{RCE} (Eq. \eqref{eq:QEr}), while at low rates it coincides with $\Eexx+R$ and is strictly larger than $\Er(R,Q)$.

To the best of our knowledge, $\Etrccclb$ is the largest known value above which the exponent distribution of  \ac{CQ} channels with arbitrary states accumulates. Moreover, we show that at rates close to the mutual information of the channel our bound $\Etrccclb$ coincides with the \ac{TRC}  and the exponent distribution concentrates around it.
\subsubsection{Second Contribution}

The second result we derive is the following: given an ensemble with rate $R$, the probability that a randomly selected code in the ensemble contains a nested code of the same rate such that each of its codewords achieves the quantum expurgated exponent approaches $1$ asymptotically as $n$ grows. Moreover, we show that such nested code can be obtained by expurgating from the original code a fraction of codewords that is vanishingly small with the codeword length. This refines Holevo's result \cite{Q_expur_general_Holevo200} that showed the existence of at least one such code, from which half the codewords need to be expurgated.\\

The starting point for both contributions is the following lemma \cite{coccoTIT2022}:
\begin{lemma}
\label{lem:root}
Let $\gamn \geq 1$ be a real-valued sequence in $n$ and let $\rv_n$ be a sequence of non-negative random variables. For any $\Grho>0$, it holds that
\beq
\PP\Bigl [ \rv_n \geq \gamn^{\Grho}\EE\left[\rv_n^\frac{1}{\Grho}\right]^{\Grho} \Bigr ]\leq \frac{1}{\gamn}.
\label{eq:0}
\eeq
\end{lemma}
Lemma \ref{lem:root} applies to arbitrary random variables and, for some families of random variables and proper choice of the sequence $\gamn$, it allows to bound from below the exponent of a randomly chosen element from the sample space with high probability as $n$ grows.

Our first contribution comes from specializing Lemma \ref{lem:root} to the error probability of an ensemble of random codes for the CQ channel. In \cite{coccoTIT2022}\footnote{Here we use a slight variation of \cite[Lemma 1]{coccoTIT2022} imposing $\gamn\geq 1$, which simplifies the proof.} it was specialized to the probability of error for pairwise independent code ensembles and generic classical channels. However, the results in \cite{coccoTIT2022} cannot be applied to  \ac{CQ} channels since they require the existence of a channel transition probability $\Wnvec$, which is not defined in the general case for CQ channels. 
In turn, it is always possible to find an equivalent \ac{CQ} channel for any given classical channel (see, e.g., \cite{sp_cq_dalai_2023} for a discussion on the subject).
This implies that the result presented in \cite{coccoTIT2022} relative to i.i.d. codes over discrete memoryless channels is a special case of what is shown in the present paper.

Our second contribution comes from specializing Lemma \ref{lem:root} to the codeword error probability and then applying a refined expurgation argument. This generalizes the  result relative to \ac{i.i.d.} codes over classical \ac{DMC} presented in \cite{coccoTIT2024_refinementExpurgation}, which can be retrieved as a special case of the result presented here for CQ channels.\\

\subsection{Code Ensembles and POVM}

A key difference with respect to  classical channels and code ensembles is that the performance of a code over a \ac{CQ} channel depends on the specific \ac{POVM} considered at the receiver. The results presented in the following consider the same \ac{POVM}s that attain the achievability results for the \ac{CQ} \ac{RCE} \cite{Q_randomCoding_Brunashev1998} and \ac{CQ} expurgated exponents \cite{Q_expur_general_Holevo200}. Specifically, the \ac{POVM} that allows to achieve the \ac{CQ} expurgated exponent is Holevo's \cite{Q_expur_general_Holevo200} tilted version of the pretty good measurement \cite{wildeBook_2021} is: 
\begin{align}\label{eq:POVM}
\Pi_j = \left(\sum_{m=1}^{M_n}(\sigma^{\x_m})^t\right)^{-\frac{1}{2}}(\sigma^{\x_j})^t \left(\sum_{m=1}^{M_n}(\sigma^{\x_m})^t\right)^{-\frac{1}{2}},
\end{align}
for $j=1,\ldots,M_n$, where $0<t<1$ is a real-valued optimization parameter. In this case the \ac{POVM} associated to a given code of the ensemble is a deterministic function of the code itself.  
One could generalize such approach by considering an ensemble of all valid \ac{POVM}s and selecting codes and \ac{POVM} independently from their respective ensemble, but will not be considered in the following.

As a final remark, the expression \eqref{eq:POVM} always exists if $\sum_{m=1}^{M_n}(\sigma^{\x_m})^t$ is full rank. 
In case the channel output alphabet contains rank-deficient element, it can happen that the sum in \eqref{eq:POVM} becomes rank deficient and \eqref{eq:POVM} does not exist. In such case the inverse $\left(\sum_{m=1}^{M_n}(\sigma^{\x_m})^t\right)^{-\frac{1}{2}}$ can be substituted with the square root of the Moore-Penrose inverse \cite{hausladen1996classical,moore1920reciprocal,bjerhammar1951application, penrose1955generalized}.

\section{First Result}\label{sec:first}
Let us define the following quantity:
\begin{align}\label{eq:ranges}
\Etrccclb=
\begin{cases}
\EexRR+R \ \ \text{ for } R \leq R^*\\
\Er(R,Q) \ \ \text{ for } R > R^*
\end{cases}
\end{align}
where $R^*=\frac{E_{\rm x}(1,Q)}{2}$, while functions $E_{\rm x}(r,Q)$, $\Eex$ and $\Er(R,Q)$ are defined in Section \ref{sec:prelim}.
Let us define the negative normalized error exponent for a random code $\Cn$ as:
\begin{align}
\En = -\frac{1}{n}\log \Pecn.
\end{align}

Our first result is the following theorem:
\begin{theorem}\label{theo:1}
For i.i.d. code ensembles over classical-quantum channels the following holds:
\begin{equation}\label{eqn:theo1_statement}
\mathbb{P}\left[ \lim_{n\rightarrow\infty}\En > \Etrccclb\right]= 1.
\end{equation}
\end{theorem}
\begin{proof}[Proof of Theorem 1]
We start with the following specialized version of Lemma \ref{lem:root}:
\begin{lemma}
\label{lem:main}
Let $\gamn \geq 1$ be a real-valued sequence in $n$. For an arbitrary random-coding ensemble and channel and any $\Grho>0$, it holds that
\beq
\PP\Bigl [ \Pecn \geq \gamn^{\Grho}\EE[\Pecn^\frac{1}{\Grho}]^{\Grho} \Bigr ]\leq \frac{1}{\gamn}.
\label{eq:4}
\eeq
\end{lemma}

Now we proceed to bound from above $\EE[\Pecn]$ and $\EE[\Pecn^\frac{1}{\Grho}]^{\Grho}$.
The first term  is the ensemble average of the error probability. An upper bound on such quantity is \cite{cheng2025errorexponentsquantumpacking}\cite{Q_expur_general_Holevo200}
\cite{renesTIT2025_tight_LB}:
\begin{align}\label{eq:pe_r}
\EE[\Pecn] \leq  \Prc,
\end{align}
where
\begin{align}
\Prc \triangleq \min_{0\leq s\leq 1} 2(M_n-1)^s\left\{\Tr \left[\left(\sum_{x\in\mathcal{X}}Q(x)\left(\sigma^x\right)^{\frac{1}{1+s}}\right)^{1+s}\right]\right\}^n.
\end{align}
The term $\EE[\Pecn^\frac{1}{\Grho}]^{\Grho}$ is the tilted ensemble-average of the error probability. An upper bound on such quantity is provided in the following lemma:\\

\begin{lemma}\label{lemma:ex}
For general \ac{CQ} channels and \ac{i.i.d.} codes and for $\Grho\geq 1$ the following holds:
\begin{align}\label{eq:lemma_ex}
\mathbb{E}\left[\Pecn^\frac{1}{\Grho} \right] \leq \Pex
\end{align}
where
\begin{align}
\Pex \triangleq \frac{1}{M_n^\frac{1}{\Grho}}M_n(M_n-1)\left(\sum_{x\in\mathcal{X}}\sum_{x'\in\mathcal{X}} Q(x)Q(x')\left( {\rm Tr}\left\{\sqrt{\sigma^{x}}\sqrt{\sigma^{x'}}\right\}\right)^\frac{1}{\Grho}\right)^n.
\end{align}
\end{lemma}
\begin{proof}[Proof of Lemma \ref{lemma:ex}]
    See Appendix.
\end{proof}
We now apply Lemma \ref{lem:main}, choosing  $\gamma_n$ to be a positive sequence such that $\gamn\geq 1$, $ \gamma_n \xrightarrow{n} \infty$ and  $\frac{1}{n}\log\gamma_n \xrightarrow{n} 0$. We also allow the parameter $\Grho$ to depend on $n$, indicating it as $\Grho_n$. Notice that Lemma \ref{lem:main}  holds $\forall n\in \mathbb{N}$. By using Equations \eqref{eq:pe_r} and \eqref{eq:lemma_ex}, we find a lower bound on the \ac{l.h.s.} of \eqref{eq:4}:
\begin{align}
\PP\Bigl [& \Pecn \geq \min_{\Grho_n\geq 1}\gamn^{\Grho_n}\EE\left[\Pecn^\frac{1}{\Grho_n}\right]^{\Grho_n} \Bigr ] \notag\\\label{eq:markov_ineq_chain_1}
=&\PP\left[ \Pecn \geq \min_{\Grho_n\geq 1}\left(\min\left\{\gamn\EE[\Pecn],\gamn^{\Grho_n}\EE\left[\Pecn^\frac{1}{\Grho_n}\right]^{\Grho_n}\right\}\right) \right]\\ 
\geq& \PP\left[ \Pecn \geq \min_{\Grho_n\geq 1}\left(\min\left\{\gamn \Prc,\gamn^{\Grho_n} \Pexn^{\Grho_n}\right\} \right)\right] \label{eq:markov_ineq_chain_2}\\
 =&\PP\left[ \Pecn \geq \min\left\{\gamn \Prc, \min_{\Grho_n\geq 1}\gamn^{\Grho_n} \Pexn^{\Grho_n}\right\} \right] \label{eq:markov_ineq_chain_3}
\end{align}
where \eqref{eq:markov_ineq_chain_1} is a rewriting of the optimization problem, \eqref{eq:markov_ineq_chain_2} follows from the bounds \eqref{eq:pe_r} and \eqref{eq:lemma_ex} on the ensemble average of the probability of error and of the tilted probability of error, respectively, while \eqref{eq:markov_ineq_chain_3} follows from the fact that the minimum over $\Grho_n$ only affects the second term within curled brackets in \eqref{eq:markov_ineq_chain_2}.
 Taking the negative normalized logarithm of the quantities inside the square brackets in \eqref{eq:markov_ineq_chain_3} and using \eqref{eq:markov_ineq_chain_1}, we obtain the following chain of inequalities:
\begin{align}\label{eq:exp_ineq_0}
 \PP\Bigl [& \Pecn \geq \min_{\Grho_n\geq 1}\gamn^{\Grho_n}\EE\Bigl[\Pecn^\frac{1}{\Grho_n}\Bigr]^{\Grho_n} \Bigr ]
 \\\notag
 \geq&\PP\left[ -\frac{1}{n}\log\Pecn \leq -\frac{1}{n}\log\left(\min\left\{\gamn \Prc, \min_{r_n\geq 1} \gamn^{{\Grho}_n} \Pexn^{\Grho_n}\right\}\right) \right] \\\notag
 \geq & \PP\left[ \En \leq \max\left\{-\frac{1}{n}\log\gamn -\frac{1}{n} +  \max_{0\leq s\leq 1}  - sR -\log \left(\Tr \left[\left(\sum_{x\in\mathcal{X}}Q(x)\left(\sigma^x\right)^{\frac{1}{1+s}}\right)^{1+s}\right]\right), \right.\right.\notag\\ \label{eq:exp_ineq_end}
 & \left.\left. 
  \max_{\Grho_n\geq 1} -{\Grho}_n\frac{\log\gamn}{n} + R  -{\Grho}_n2R -{\Grho}_n\log\left(\sum_{x\in\mathcal{X}}\sum_{x'\in\mathcal{X}} Q(x)Q(x')\left( {\rm Tr}\left\{\sqrt{\sigma^{x}}\sqrt{\sigma^{x'}}\right\}\right)^\frac{1}{{\Grho}_n}\right)\right\} \right] \\\label{eq:exp_ineq_end_p1}
 = &\PP\biggl[ \En \leq \max\left\{\iota_n+\Er(R,Q),\biggr.\right.\\\notag
 &\left.\biggl.\max_{\Grho_n\geq 1}\  \delta_n(\Grho_n) + R - \Grho_n 2R + \Exqn \right\} \biggr]
\end{align}
where in \eqref{eq:exp_ineq_end} we used the facts that $-\frac{1}{n}\log (M_n-1) > -R$ and $-\frac{1}{n}\log M_n(M_n-1) > -2R$, while in \eqref{eq:exp_ineq_end_p1}  we rewrote the \ac{r.h.s.} of the inequality in terms of $\Er(R,Q)$ and $\Exq$, given in \eqref{eq:QEr} and \eqref{eqn:ex}, respectively, and we defined
$\iota_n = -\frac{(1+\log \gamma_n)}{n}$ and $\delta_n(\Grho_n) = -{\Grho}_n\frac{\log \gamma_n}{n}$.
Using \eqref{eq:exp_ineq_end_p1} in \eqref{eq:4} we have:
\begin{align}\label{eq:exp_ineq}
\PP\left[\right.&\left.\En \leq \max\left\{\iota_n+\Er(R,Q),\delta_n(\hat{\Grho}_n) +R  - \hat{\Grho}_n 2R + \Exqnh \right\} \right] \\\label{eq:exp_ineq_b}
\leq &\PP\Bigl [ \Pecn \geq \min_{\Grho_n\geq 1}\gamn^{\Grho_n}\EE\Bigl[\Pecn^\frac{1}{\Grho_n}\Bigr]^{\Grho_n} \Bigr ]\\\label{eq:exp_ineq_c}
\leq& \frac{1}{\gamma_n}
\end{align}
where in \eqref{eq:exp_ineq} we defined:
\begin{align}
    \hat{\Grho}_n = \arg\max_{\Grho_n\geq 1} \delta_n(\Grho_n) + R - \Grho_n 2R + \Exqn,
\end{align}
that is, $\hat{\Grho}_n$ is the parameter that maximizes the expression for a given $n$. Notice that the only term that actually depends on $n$ in the second line of expression \eqref{eq:exp_ineq_end_p1} is $\delta_n(\Grho_n)$.
Removing $\delta_n(\Grho_n)$ from such expression, we define the maximizing parameter as:
\begin{align}\label{eqn:r_hat}
    \hat{\Grho} = \arg\max_{\Grho\geq 1} R - \Grho 2R + \Exq,
\end{align}
which does not depend on $n$.

Since $\gamma_n$  in \eqref{eq:exp_ineq}-\eqref{eq:exp_ineq_c} is chosen so that it goes to infinity sub-exponentially, we have the following implications: the first is that \eqref{eq:exp_ineq_c} goes to $0$ as $n\rightarrow\infty$. The second is that $\iota_n\xrightarrow{n}0$. The third implication is that for all rates, ensembles and codes for which $\hat{\Grho} <\infty$ in \eqref{eqn:r_hat}, we have that  $\hat{\Grho}_n\xrightarrow{n} \hat{\Grho}$, since $-\frac{\log \gamn}{n}\xrightarrow{n}0$, which implies $\delta_n(\hat{\Grho}_n)=-\hat{\Grho}_n\frac{\log \gamn}{n}\xrightarrow{n} 0$. In turn, these facts yield:
\begin{align}\notag
\lim_{n\rightarrow\infty}\delta_n(\hat{\Grho}_n) + R - \hat{\Grho}_n 2R +\Exqnh  = \EexRR+R.
\end{align}
For some CQ channels, code ensembles and rates, it can happen that $\hat{\Grho}\rightarrow\infty$. Such configurations must be addressed to ensure that Theorem \ref{theo:1} does not yield a trivial result. We distinguish two cases, namely, zero rate and strictly positive rates.
\subsection{$R=0$ and $\hat{\Grho}\rightarrow\infty$}\label{sec:R0_r_inf}
A rate equal to $0$ occurs if the number of codewords $M_n$ grows sub-exponentially or is constant in $n$, which implies $R=\frac{\log M_n}{n}\rightarrow 0$.

The behavior of $\delta_n(\hat{\Grho}_n)$ for $R=0$ is considered in the following lemma:
\begin{lemma}\label{lemma:R0}
    For general \ac{CQ} channel and \ac{i.i.d.} codes, if $\frac{\log M_n}{n}\rightarrow 0$ the following holds:
\begin{align}
    \lim_{n\rightarrow\infty} \delta_n(\hat{\Grho}_n) = 0.
\end{align}
\end{lemma}
\begin{proof}[Proof of Lemma \ref{lemma:R0}]
    See Appendix.
\end{proof}
\subsection{$R>0$ and $\hat{\Grho}\rightarrow\infty$}
The second term in Eqn. \eqref{eq:exp_ineq_end_p1} can be written as:
\begin{align}\label{eq:rect}
     R (1- 2\Grho) + \Exq - \Grho\frac{\log \gamn}{n},
\end{align}
which, for any value of the optimization parameter $r$, is linear in $R$ with slope $(1- 2\Grho)<0$. We will now calculate the $R$-axis intercept of \eqref{eq:rect} and then take the limit for ${\Grho}\rightarrow\infty$ first (which corresponds to \eqref{eq:rect} having an asymptotically infinite slope), and  for $n\rightarrow\infty$ afterwards\footnote{Note that the order in which the limits are taken has no impact in this case, that is, the limit does not change if the order is inverted.}. This gives us the rate at which \eqref{eq:rect} has a vertical asymptote. Below such rate, which we indicate with $\Rinf$, the quantity \eqref{eq:rect} (and thus the maximization at the second term in Eqn. \eqref{eq:exp_ineq_end_p1}) diverges to $\infty$, as described in the following.  
Setting \eqref{eq:rect} to $0$, solving in $R$ and taking the limit for ${\Grho}\rightarrow\infty$ we have:
\begin{align}\label{eq:rect2}
    \lim_{{\Grho}\rightarrow\infty} &\frac{\Exq}{2{\Grho}-1} - \frac{{\Grho}}{2{\Grho}-1}\frac{\log \gamn}{n}=-\frac{1}{2}\log \PP\left[{\rm Tr}\left\{\sqrt{\sigma^{x}}\sqrt{\sigma^{x'}}\right\}>0\right] - \frac{{1}}{2}\frac{\log \gamn}{n},
\end{align}
where the term $\log \PP\left[{\rm Tr}\left\{\sqrt{\sigma^{x}}\sqrt{\sigma^{x'}}\right\}>0\right]$ comes from applying L'H\^opital's rule.
Finally, taking the limit for $n\rightarrow\infty$ we find:
\begin{align}\label{eq:rect3}
    \Rinf =-\frac{1}{2}\log \PP\left[{\rm Tr}\left\{\sqrt{\sigma^{x}}\sqrt{\sigma^{x'}}\right\}>0\right].
\end{align}
Notice that the term ${\rm Tr}\left\{\sqrt{\sigma^{x}}\sqrt{\sigma^{x'}}\right\}$ in the equation above is always nonzero unless states $\sigma^{x}$ and $\sigma^{x'}$ are orthogonal. Thus, expression \eqref{eq:rect3} can only be non-zero if at least two states are pairwise orthogonal.
Let us now consider the following facts: i) expression \eqref{eq:rect} has a vertical asymptote in the point $R=\Rinf$, ii) expression \eqref{eq:rect}, seen as a function of $R$, has a negative slope, 
iii) $\hat{\Grho}$ in \eqref{eqn:r_hat} is nonincreasing in $R$ \cite{Q_expur_general_Holevo200}. These three facts imply that, for  $R<\Rinf$, the maximum over $\Grho$ of \eqref{eq:rect} takes an infinite value, which is achieved for an asymptotically large ${\Grho}$. The threshold rate $\Rinf$ can be equal to $0$ or can be positive. The procedure just presented is similar to the derivation of Gallager's lower bound  on the zero-error capacity \cite{gallagerBook} for the expurgated exponent in classical channels. 
 In a similar way, the threshold rate $\Rinf$ given in \eqref{eq:rect3} is related to the zero-error capacity of \ac{CQ} channels, for which $2\Rinf$ is a lower bound. A thorough study of the zero-error capacity for \ac{CQ} channels is out of the scope of the present paper. A discussion and other references on the topic can be found in, e.g., \cite{sp_cq_dalai_2023} and \cite{ze_cap_duan_TIT2013}.\\
The case $R\geq\Rinf$ is covered in the discussion before subsection \eqref{sec:R0_r_inf}.

The analysis just carried out yields the following: 
\begin{enumerate}
\item $\delta_n(\hat{\Grho}_n)\rightarrow 0$ $\forall R\geq\Rinf$ for any \ac{CQ} channel, which implies that the \ac{r.h.s.} of the inequality in \eqref{eq:exp_ineq_end_p1} is strictly positive.
\item $R-\Grho_n2R+\Exqn + \delta_n(\Grho_n) \rightarrow \infty$ for $0<R<\Rinf$, which implies that the \ac{r.h.s.} of the inequality in \eqref{eq:exp_ineq_end_p1} is $\infty$.\\
\end{enumerate}

Let us now take the limit for $n\rightarrow\infty$ in \eqref{eq:exp_ineq} and \eqref{eq:exp_ineq_c}. This yields:
\begin{align}\label{eq:exp_ineq_2}
\lim_{n\rightarrow \infty}\PP\left[ \En \leq \max\left\{\iota_n+\Er(R,Q),\delta_n(\hat{\Grho}_n) + R -\hat{\Grho}_n2R + \Exqnh \right\} \right] \leq \lim_{n\rightarrow \infty}\frac{1}{\gamma_n}.
\end{align}
In order to bring the limit inside the probability at the \ac{l.h.s.} of \eqref{eq:exp_ineq_2} we apply the Borell-Cantelli lemma \cite[Sec. 2.3]{Durett_probab_Book}. To do so, the sequence $\gamma_n$ has to satisfy the following two conditions:
\begin{align}
\gamma_n&\rightarrow \infty\\
\sum_{n=1}^{\infty}\frac{1}{\gamma_n}&<\infty.
\end{align}
The first one is satisfied by imposing that $\gamn$ diverges. The second condition is satisfied by choosing, for instance, $\gamn=n^{a}$ with $a>1$. This guarantees the sub-exponential growth of $\gamn$ which is required for $\iota_n$ and $\delta_n(\hat{\Grho}_n)$ to vanish.
Using such a sequence we can bring the limit inside the probability in \eqref{eq:exp_ineq_2} and obtain:
\begin{align}
&\PP\left[\lim_{n\rightarrow \infty} \En \leq \lim_{n\rightarrow \infty} \max\left\{\iota_n+\Er(R,Q),\delta_n(\hat{\Grho}_n) + R - \hat{\Grho}_n2R + \Exqnh \right\} \right] \notag\\
&= \PP\left[\lim_{n\rightarrow \infty} \En \leq  \max\left\{\Er(R,Q), \EexRR + R \right\} \right]\label{eq:exp_ineq_3a}.
\end{align}
The result of  $\max\left\{\Er(R,Q), \EexRR + R\right\}$ depends on the rate. By studying the two functions involved (see expressions \eqref{eq:QEr} and \eqref{eqn:expu}) it can be shown that\footnote{We do not report the derivation here since it is lengthy and is similar in spirit to the one used in \cite[Appendix]{Q_expur_general_Holevo200}.}:
 \begin{align}\label{eq:ranges_2}
\max\left\{\Er(R,Q), \right.\left.\EexRR + R \right\}=\begin{cases}
\EexRR+R \ \ \text{ for } R \leq R^*\\
\Er(R,Q) \ \ \text{ for } R > R^*
\end{cases}
\end{align}
where $R^*=\frac{1}{2}\frac{\partial E_{\rm x}(r,Q)}{\partial r}|_{r=1}$. By combining \eqref{eq:exp_ineq_2} and \eqref{eq:exp_ineq_3a} and using \eqref{eq:ranges_2}, for general \ac{CQ} channels and any rate we have:
\begin{align}
&= \PP\left[\lim_{n\rightarrow \infty} \En \leq  \max\left\{\Er(R,Q), \EexRR + R \right\} \right]\notag\\
&=\mathbb{P}\left[ \lim_{n\rightarrow\infty}\En \leq \Etrccclb\right]\label{eq:exp_ineq_4}\notag\\
& \leq \lim_{n\rightarrow \infty}\frac{1}{\gamma_n}\notag\\
&= 0\notag.
\end{align}
 This concludes the proof.
\end{proof}

In general $\Etrccclb\leq \Etrccc$, where $\Etrccc=\lim_{n\rightarrow\infty}\EE[\En]$ is the ensemble-average exponent of the error probability, i.e., the \ac{TRC} exponent \cite{merhav_TIT2018}.
This follows from Lemma \ref{lem:main} and the fact that:
\begin{align}
&\lim_{n\rightarrow\infty}\max_{r_n\in[1,\infty)}-\frac{1}{n}\log\gamn^{\Grho_n}\EE\left[\Pecn^\frac{1}{\Grho_n}\right]^{\Grho_n} \\
 &\leq \lim_{n\rightarrow\infty}\max_{r_n\in[1,\infty)}-\frac{1}{n}\log\EE\left[\Pecn^\frac{1}{\Grho_n}\right]^{\Grho_n}\label{eq:ineq_lb1}
\\ 
&= \lim_{n\rightarrow\infty}\lim_{r\rightarrow\infty} -\frac{1}{n}\log\EE\left[\Pecn^\frac{1}{\Grho}\right]^{\Grho}\label{eq:ineq_lb2}  \\ 
&=\lim_{n\rightarrow\infty}\EE[\En].\label{eq:ineq_lb3}
\end{align}
where \eqref{eq:ineq_lb1} holds since $\gamma_n\geq 1$, \eqref{eq:ineq_lb2} is because $-\frac{1}{n}\log\EE\left[\Pecn^\frac{1}{\Grho}\right]^{\Grho}$ is monotonically increasing in $\Grho$, while \eqref{eq:ineq_lb3} follows from solving the limit in $\Grho$.

\subsection{Concentration at the Exponent at High Rates}
Theorem \ref{theo:1} shows that the vast majority of codes in \ac{i.i.d.} ensembles over \ac{CQ} channels have an exponent at least as large as $\Etrccclb$. 
This has the following implication. The works \cite{sp_cq_dalai_2023} and \cite{renesTIT2025_tight_LB} showed that at rates close to capacity (i.e., larger than the cutoff rate, as defined in \cite[Chapter 5]{gallagerBook}) $\Er(Q)$ is, once optimized in the input distribution, the largest possible exponent in a random code ensemble for $\ac{CQ}$ channels with generic output states. Our Theorem \ref{theo:1} states that the probability of finding a code in the ensemble having an exponent smaller than such quantity goes to $0$ asymptotically in $n$. Combining the two results, it follows that the exponent probability distribution across the ensemble concentrates around $\Er(R)$ in the high-rate regime, since no exponent larger than $\Er(R)$ exists in the ensemble and the vast majority of codes achieve $\Er(R)$. This also implies that, at rates close to the mutual information of the channel, $\Er(R)$ coincides with the \ac{TRC} exponent $\Etrccc$

\section{Second Result}\label{sec:refexp}
Our second contribution strengthens Holevo's result on expurgation for CQ channels with arbitrary states. We show that, for an arbitrarily small $\epsilon>0$, the probability of finding a code with $M_n'=(1+\epsilon)M_n$ codewords that contains a nested code with at least $M_n$ codewords, each of which achieving the CQ expurgated exponent, tends to $1$ with the codeword length $n$.

\subsection{Codeword error exponents}
For a given code $\Cnalldet$ and a POVM $\{\Pi_0,\Pi_1,\ldots,\Pi_{M_n'}\}$, the probability of error $\Pecnmalldet$ conditioned to transmitting codeword $m$ is given in \eqref{eq:pe_m}. Let us define the corresponding codeword error exponent as
\begin{align}\label{eq:cwExp}
E_m(\Cnalldet) \triangleq -\frac{1}{n}\log \Pecnmalldet.
\end{align}
If a random code ensemble is considered, a random variable can be defined according to \eqref{eq:cwExp} as follows:
\begin{align}
E_m(\Cnall) \triangleq -\frac{1}{n}\log \Pecnmall.
\end{align}

Before moving to the main result of this section, we need to introduce the following bound on the lower tail of the codeword exponent distribution:
\begin{lemma}\label{lem:codeword_tail}
For general CQ channels with arbitrary output states, i.i.d.\ code ensembles, any $m\in\{1,\ldots,M_n'\}$, any sequence $\gamma_n\ge 1$, and any $\GrhoEx\ge 1$,
\begin{align}\label{eq:codeword_tail_ex}
\mathbb{P}\!\left[
E_m(\Cnall) > E_{\rm x}(\GrhoEx,Q)-\GrhoEx R - \frac{\GrhoEx\log \gamma_n}{n} \right]> 1- \frac{1}{\gamma_n}.
\end{align}

\end{lemma}

\begin{proof}
The proof proceeds by specializing Lemma~\ref{lem:root} to the random variable $\Pecnmall$:
\begin{align}\label{eq:markov_codeword}
\mathbb{P}\!\left[\Pecnmall \ge \gamma_n^{\GrhoEx}\,\mathbb{E}\!\left[\Pecnmall^{\frac{1}{\GrhoEx}}\right]^{\GrhoEx}\right]\le \frac{1}{\gamma_n}
\end{align}
and proceeding in a similar way as in Section \ref{sec:first}, keeping in mind the fact that now we calculate a bound on the codeword error probability rather than on the error probability of the whole code.
\end{proof}

Let us indicate with $\hat{\GrhoEx}$ the optimization parameter yielding the CQ expurgated exponent, that is:
\begin{align}\label{eqn:r_hatEx}
    \hat{\GrhoEx} = \arg\max_{\GrhoEx\geq 1} \left\{E_{\rm x}(\GrhoEx,Q)-\GrhoEx R\right\}.
\end{align}
Notice that the optimal parameter in \eqref{eqn:r_hatEx} is not necessarily the same as the one given in the optimization \eqref{eqn:r_hat} due to the factor $2$ which multiplies the rate on the former expression. Hence,  a different notation has been adopted to avoid confusion.

Let us now define the following indicator function for a given code $\Cnalldet$:
\begin{align}\label{eq:indi}
\varphi_m(\Cnalldet) \triangleq \mathbf{1}\!\left\{E_m(\Cnalldet) > E_{\rm ex}(R,Q)-\deltaEx_n(\hat{\GrhoEx}) \right\},
\end{align}
where $\deltaEx_n(\hat{\GrhoEx}) = \hat{\GrhoEx}\frac{\log \gamma_n}{n}$ is a positive sequence in $n$.
The function $\varphi_m(\Cnalldet)$ is $1$ whenever the $m$-th codeword of code $\Cnalldet$ attains an exponent larger than $E_{\rm ex}(R,Q)-\deltaEx_n(\hat{\GrhoEx})$. Notice that $\deltaEx_n(\hat{\GrhoEx})$ can be studied using the same approach as for $\delta_n$ in Section \ref{sec:first}, with similar results. In other words, for all channels and code ensembles considered in the present work, either $\deltaEx_n(\hat{\GrhoEx})\xrightarrow{n} 0$ or it diverges at a slower rate with respect to $ E_{\rm ex}(R,Q)$, so that the right-hand-side of the inequality in \eqref{eq:indi} always converges to $E_{\rm ex}(R,Q)$.
We indicate with $\Phi(\Cnalldet)$ the number of codewords in the code $\Cnalldet$ that satisfy the condition in \eqref{eq:indi}, 
that is:
\begin{align}
\Phi(\Cnalldet)\triangleq \sum_{m=1}^{M_n'}\varphi_m(\Cnalldet).
\end{align}

The main result of this section is given by the following theorem.
\begin{theorem}\label{theo:refexp_cq}
Consider an \ac{i.i.d.} CQ code ensemble with arbitrary output states and $M_n'=(1+\epsilon)M_n$ codewords for some $\epsilon>0$ \footnote{For the sake of correctness, the quantity $(1+\epsilon)M_n$ may not be an integer number and, with a slight abuse of notation, has to be interpreted as $\lceil(1+\epsilon)M_n\rceil$. This does not hinder the validity of the results and avoids cluttering the notation.} . 
Then the following holds:
\begin{align}\label{eq:refexp_statement}
\lim_{n\to\infty}\mathbb{P}\!\left[\Phi(\Cnall)>M_n\right]=1.
\end{align}
\end{theorem}

The theorem states that, for all code ensembles with $M_n'=(1+\epsilon)M_n$ codewords over any CQ channel, with probability approaching $1$ asymptotically in $n$, a randomly selected code can be expurgated to a smaller code of the same rate with $M_n$ codewords such that each codeword achieves the CQ expurgated exponent. This result is quite general, in that it holds for all CQ channels, ensembles and rates, which is also the case for the results presented in Section \ref{sec:first}.

\begin{proof} Let $\gamma_n\to\infty$ with $\frac{\log \gamma_n}{n}\to 0$. 
From Lemma~\ref{lem:codeword_tail}, for every $m$ we have $\mathbb{E}[\varphi_m(\Cnall)]> 1-\frac{1}{\gamma_n}$. Hence
\begin{align}\label{eq:psi}
\mathbb{E}[\Phi(\Cnall)] > (1-\tfrac{1}{\gamma_n})M_n'.
\end{align}
Let us indicate with $\Psi(\Cnalldet)\triangleq M_n'-\Phi(\Cnalldet)$  the number of codewords not achieving the expurgated exponent in a given code $\Cnalldet$. From Eqn. \eqref{eq:psi} it follows that:
\begin{align}
\mathbb{E}[\Psi(\Cnall)] \le \frac{M_n'}{\gamma_n}.
\end{align}
By Markov's inequality and for sufficiently large $n$ we have:
\begin{align}\label{eq:psi_markov}
\mathbb{P}\!\left[\Psi(\Cnall)\geq\frac{M_n'}{\sqrt{\gamma_n}}\right]\le \frac{1}{\sqrt{\gamma_n}}.
\end{align}
To prove our main result, we write the tail probability in~\eqref{eq:refexp_statement} using \eqref{eq:psi_markov}: 
\begin{align}\label{eq:chain_good1}
&\PP\Big[\Phi(\Cnall)> M_n\Big]\notag\\&\
= 1-\PP\Big[\Phi(\Cnall) \leq M_n\Big]\\
&\ = 1-\PP\Big[\Psi(\Cnall)> M_n'-M_n\Big]\\
&\ = 1-\PP\Big[\Psi(\Cnall)> \epsilon M_n\Big],
\end{align}
where we used the definitions of $\Psi(\Cnall)$ and $M_n'$. Since $\gamma_n$ tends to infinity, there must exist an $n_0\in\mathbb{N}$ such that $\epsilon > \frac{1+\epsilon}{\sqrt{\gamma_n}}$ for $n>n_0$ and therefore
\begin{align}
&\lim_{n\to\infty}\PP\Big[\Phi(\Cnall) > M_n\Big]\notag\\ 
&\ =\lim_{n\to\infty}  1-\PP\Big[\Psi(\Cnall)> \epsilon M_n\Big]\\
&\ \geq \lim_{n\to\infty} 1-\PP\Big[\Psi(\Cnall) \geq \frac{(1+\epsilon)M_n}{\sqrt{\gamma_n}}\Big]\label{eq:chain_good1_a}\\
&\ \geq  \lim_{n\to\infty} 1-\frac{1}{\sqrt{\gamn}}, \label{eq:chain_good1_b}\\
&\ =1
\end{align}
where \eqref{eq:chain_good1_b} follows from \eqref{eq:psi_markov}. This concludes the proof.
\end{proof}

\subsection{Note on the POVM}
While formally similar, a key difference between the result just presented and its classical counterpart \cite{coccoTIT2024_refinementExpurgation} is that the codeword error probability $\Pecnmalldet$ for each given code of size $M'_n$ depends on the specific \ac{POVM} that has been chosen. Specifically, as indicated in \eqref{eq:POVM} the chosen \ac{POVM} is, for $j=1,\ldots, M'_n$:
\begin{align}\label{eq:POVM_M1}
\Pi_j = \left(\sum_{m=1}^{M_n'}(\sigma^{\x_m})^t\right)^{-\frac{1}{2}}(\sigma^{\x_j})^t \left(\sum_{m=1}^{M_n'}(\sigma^{\x_m})^t\right)^{-\frac{1}{2}},
\end{align}
and it holds that:
\begin{align}
\sum_{j=0}^{M'_n}\Pi_j=I,
\end{align}
where
\begin{align}
\Pi_0=I - \sum_{j=1}^{M'_n}\Pi_j. 
\end{align}
By Lemma \ref{lem:codeword_tail}, this \ac{POVM} guarantees the achievability of $E_{\rm ex}(R,Q)$ for most codewords in most codes of the ensemble. After expurgating the size-$M_n'$ code by removing  $\epsilon M_n$ codewords, we obtain a code of size $M_n$ and a new \ac{POVM} has to be chosen. While this is implicitly assumed in \cite{Q_expur_general_Holevo200}, here we discuss this point for the sake of completeness. In order to still guarantee the achievability of the expurgated exponent for each codeword, the POVM for the expurgated code should include all elements of the original \ac{POVM} \eqref{eq:POVM_M1} relative to the codewords that were expurgated. Formally, assume without loss of generality that the last $\epsilon M_n$ codewords were expurgated from $\Cnalldet$ and let $\Pi_j^{ex}$ be the $j$-th element of the new \ac{POVM}. Then, for $j=1\ldots, M_n$:
\begin{align}
\Pi_j^{ex} = \Pi_j,
\end{align}
and 
\begin{align}
\Pi_0^{ex}&=I - \sum_{j=1}^{M_n}\Pi_j^{ex},\\
&= I + \sum_{j=M_n+1}^{M'_n}\Pi_j - \sum_{j=1}^{M'_n}\Pi_j,
\end{align}
that is, the elements of the old \ac{POVM} relative to the expurgated codewords are absorbed by $\Pi_0^{ex}$. Notice that this change in the \ac{POVM} does not alter the codeword error probability, since, for $j=1\ldots, M_n$:
\begin{align}
\PecnjallMdet = 1- \tr\{\Pi_{j}^{ex}\sigma^{\x_j} \}\\
= 1- \tr\{\Pi_{j}\sigma^{\x_j} \}.
\end{align}

\section{Numerical Results}
In this section we numerically evaluate the bound derived in Theorem \ref{theo:1} and analyze its dependence on relevant channel parameters. For completeness, the expurgated exponent expression associated with Theorem \ref{theo:refexp_cq} is also evaluated. To account for both pure and mixed states, we describe the system using density matrices. Specifically, without loss of generality, we adopt a Pauli basis expansion parametrized by the state purity $\Pur=\text{Tr}\{(\sigma^x)^2\}$ \cite{wildeBook_2021}, where $\sigma^x$ denotes the density matrix of the state. Beyond its intrinsic theoretical importance, considering mixed states in \ac{CQ} channels is practically motivated by environmental interactions that cause partial decoherence in otherwise pure states, such as amplitude damping or depolarizing channels \cite{DATTA2018243}. Crucially, this mathematical formalism universally accommodates any quantum channel configuration. While it can be used to model a preparation channel to capture imperfections during state initialization, it equally applies to scenarios where a state passes through a quantum-quantum channel after preparation, including applications not affected by preparation noise.

We consider a quaternary \ac{CQ} channel with a symmetric input distribution, that is, $Q(x)=1/4$ $\forall x\in\mathcal{X}$. Since the distribution is fixed, in the remaining part of this section we remove the dependency on $Q$ from all expressions. Let $\Pur$ be the purity of the output state $\sigma^x$. We consider two different channels, one having all mixed states and the other having two mixed states and two pure, mutually orthogonal states. 

\subsection{All Mixed States}
\begin{figure}[htbp] 
 \begin{center} \includegraphics[width=.7\linewidth,draft=false]{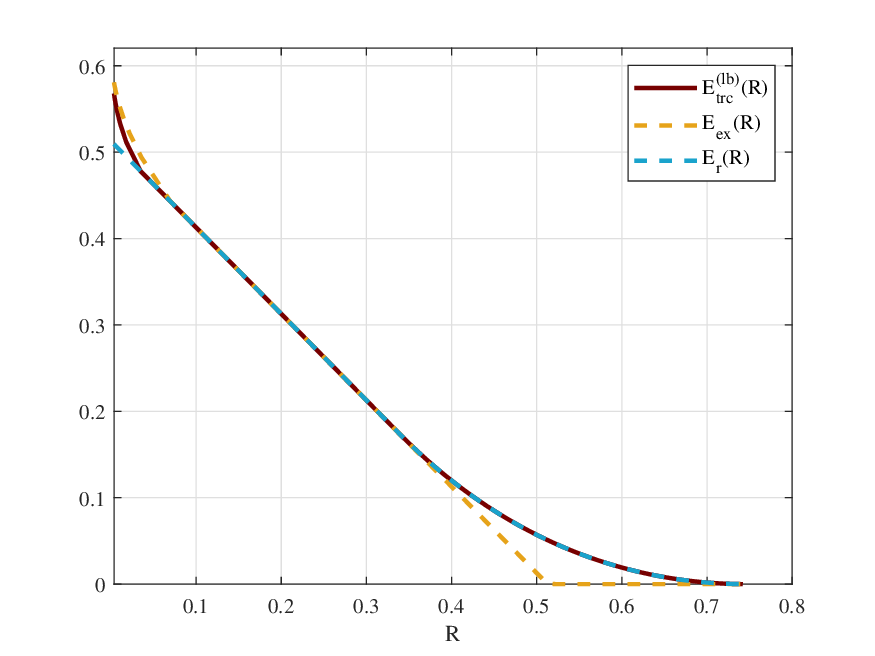}
 \caption{Exponent $\EtrccclbNoQ$ versus rate $R$ for a \ac{CQ} channel with quaternary symmetric input distribution and purity $\Pur=0.95$. The \ac{RCE} $\Er(R)$ and the expurgated exponent $\EexNoQ$ are also shown for comparison. For the considered setup $C\simeq 0.745$, $R^*\simeq 0.051$, $R_{\infty}=0$.}\label{fig:Rinf_zero}
 \end{center}
 \end{figure}

For a given purity $\Pur$, assumed to be the same for $x=1,2,3,4$, the chosen output states can be expressed in the Pauli basis as: 
\begin{align}
    \sigma^1 = \frac{1}{2}\left( I + A\sin(\theta)P_X + A\cos(\theta)P_Z\right)\\
    \sigma^2 = \frac{1}{2}\left( I + A\sin(\theta)P_X - A\cos(\theta)P_Z\right)\\
    \sigma^3 = \frac{1}{2}\left( I + A\sin(\theta)P_Y + A\cos(\theta)P_Z\right)\\
    \sigma^4 = \frac{1}{2}\left( I + A\sin(\theta)P_Y - A\cos(\theta)P_Z\right)
\end{align}
where $A=\sqrt{2\Pur - 1}$, $\theta=\frac{\pi}{6}$ unless otherwise specified, $I$ is the $2\times 2$ identity matrix while $P_X$, $P_Y$ and $P_Z$ are the Pauli matrices. Notice that the coefficient relative to $P_X$ is set to zero for states $\sigma^3,\sigma^4$ while the coefficient relative to $P_Y$ is set to zero for states $\sigma^1,\sigma^2$.
In Fig. \ref{fig:Rinf_zero} $\EtrccclbNoQ$ is plotted versus the rate for a \ac{CQ} channel with the input alphabet just described and $\Pur=0.95$. The \ac{RCE} $\Er(R)$ and the expurgated exponent are also shown. It can be seen that $\EtrccclbNoQ>\Er(R)$ for $R<R^*\triangleq\frac{E_{\rm x}(1)}{2}\simeq 0.051$. Both exponents are zero for rates larger than the capacity $C\simeq 0.745$. Since all states are mixed and a two-dimensional Hilbert space is considered, according to \eqref{eq:rect3} we have $R_{\infty}=0$.
 
 \subsection{Pure and Mixed States}
\begin{figure}[htbp] 
 \begin{center} \includegraphics[width=.7\linewidth,draft=false]{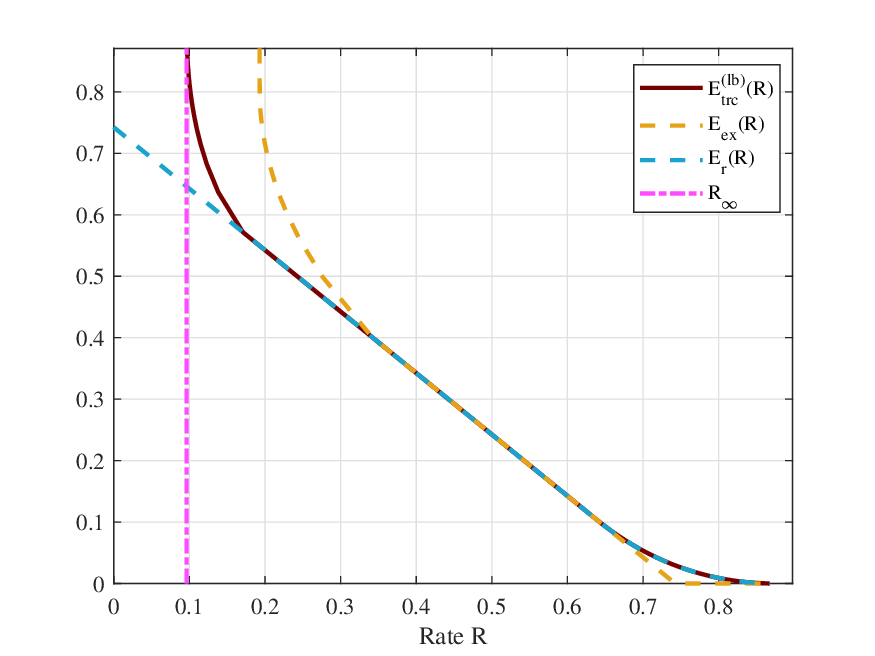}
 \caption{Exponent $\EtrccclbNoQ$ versus rate $R$ for a \ac{CQ} channel with quaternary symmetric input distribution and purity $\Pur=0.95$. The \ac{RCE} $\Er(R)$ and the expurgated exponent $\EexNoQ$ are also shown for comparison. For the considered setup $C\simeq 0.873$, $R^*\simeq 0.199$, $R_{\infty}=0.096$. The latter is represented as a vertical dash-dotted line.}\label{fig:Rinf_positive}
 \end{center}
 \end{figure}

In this setup two of the output alphabet states are mixed while the other two are pure and pairwise orthogonal. Their description in the Pauli basis is:
\begin{align}
    \sigma^1 &= \frac{1}{2}\left( I + A\sin(\theta)P_X + A\cos(\theta)P_Z\right)\\
    \sigma^2 &= \frac{1}{2}\left( I + A\sin(\theta)P_X - A\cos(\theta)P_Z\right)\\
    \sigma^3 &= \frac{1}{2}\left( I + P_Z\right)\\
    \sigma^4 &= \frac{1}{2}\left( I  - P_Z\right).
\end{align}
In Fig. \ref{fig:Rinf_positive} $\EtrccclbNoQ$ is plotted versus the rate for the considered \ac{CQ} channel with purity $\Pur=0.95$ in $\sigma^1$ and $\sigma^2$. The \ac{RCE} $\Er(R)$ and the expurgated exponent are also shown. It can be seen that $\EtrccclbNoQ>\Er(R)$ for $R<R^*\simeq 0.199$. Both exponents are zero for rates larger than the mutual information of the channel $C\simeq 0.873$. Since states $\sigma^3$ and $\sigma^4$ are pairwise orthogonal, the term ${\rm Tr}\left\{\sqrt{\sigma^{x}}\sqrt{\sigma^{x'}}\right\}$ in \eqref{eq:rect3} is zero with a strictly positive probability and $R_{\infty}=0.096$. It can be seen that, as proven in Section \ref{sec:first}, in correspondence of $R_{\infty}$ (which is represented as a vertical dash-dotted line) the plot has a vertical asymptote, and	 for $R<R_{\infty}$ the exponent $\EtrccclbNoQ$ takes an infinite value. Similarly, for all rates in the region $R<2R_{\infty}$, the expurgated exponent diverges.  This is due to the fact that, since at least two of the states ($\sigma^3$ and $\sigma^4$) cannot be confused, it is possible to find codes such that, at low rates, the probability of error is exactly zero. Our first result implies that, in fact, for $R<R_{\infty}$, most codes in the ensemble have such property. As a side remark, notice that, for this setup, the chosen uniform probability distribution is not the optimal one due to the channel asymmetry, and has been chosen only for an illustrative purpose. Optimizing on the input distribution might lead to larger error exponents in the rate region in which they are finite.
 
 \subsection{Binary Channels}
In the following we consider a different setup, specifically, we consider a binary \ac{CQ} channel with a symmetric input distribution, that is, $Q(1)=Q(2)=1/2$, and we show the impact of different channel parameters on the exponents. The output states can be expressed in the Pauli basis as: 
\begin{align}
    \sigma^1 = \frac{1}{2}\left( I + A\sin(\theta)P_X + A\cos(\theta)P_Z\right)\\
    \sigma^2 = \frac{1}{2}\left( I + A\sin(\theta)P_X - A\cos(\theta)P_Z\right).
\end{align}
 \begin{figure}[htbp] 
 \begin{center}
\includegraphics[width=.7\linewidth,draft=false]{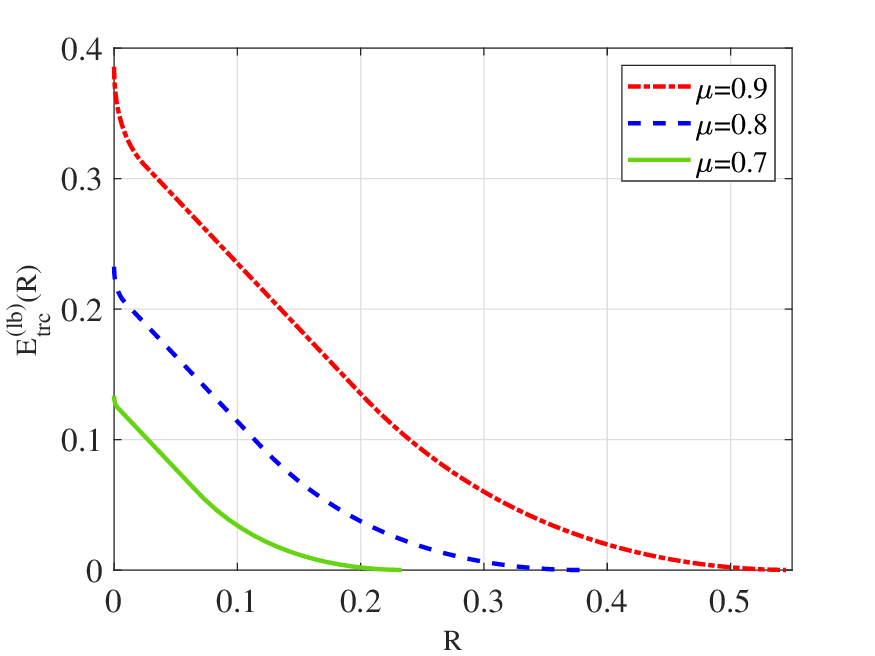}
 \caption{Exponent $\EtrccclbNoQ$ plotted versus $R$ for a \ac{CQ} channel with symmetric binary input distribution and different values of purity $\Pur$.}\label{fig:purities}
 \end{center}
 \end{figure}
 In Fig. \ref{fig:purities} $\EtrccclbNoQ$ is plotted versus $R$ for a symmetric binary input and different values of purity $\Pur$. For a given rate, the exponent $\EtrccclbNoQ$ becomes smaller as the purity decreases. Also the rate $R^*$ below which the exponent $\EtrccclbNoQ$ is strictly larger than $\Er(R)$ shows the same trend, going from $0.025$ for $\Pur=0.9$ to $0.003$ for $\Pur=0.7$.
  \begin{figure}[htbp!] 
 \begin{center}
\includegraphics[width=.7\linewidth,draft=false]{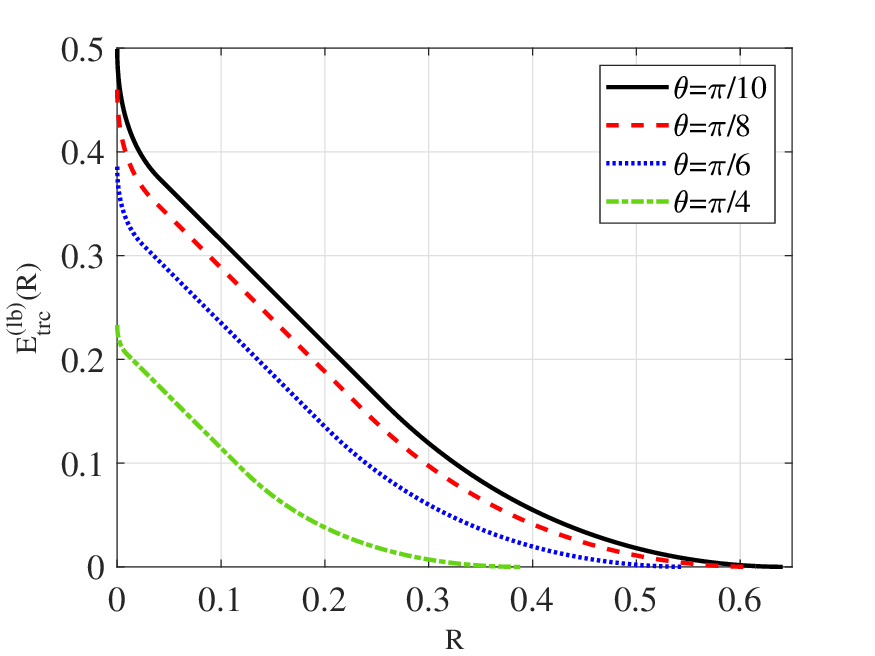}
 \caption{Exponent $\EtrccclbNoQ$ plotted versus $R$ for a \ac{CQ} channel with symmetric binary input distribution and different values of the angle $\theta$.}\label{fig:theta}
 \end{center}
 \end{figure}
In Fig. \ref{fig:theta} we plot $\Etrccclb$ versus $R$ for a symmetric input distribution, $\mu=0.9$ and different values of the angle $\theta$. It can be seen how a larger angle corresponds to a smaller exponent for a given rate. This is because the probability of error for the measurements on the output states increases. In the limiting case of pure states ($A=1$) with $\theta=0$ one obtains orthogonal output states, for which the probability of error is exactly zero.

\section{Conclusion}
In this work we have shown that for any \ac{i.i.d.} ensemble of codes over \ac{CQ} channels, the mass of the exponent distribution accumulates above the threshold $\Etrccclb$. At rates close to the channel's mutual information, $\Etrccclb$ coincides with the \ac{RCE}, while at low rates it matches $\Eexx + R$ and remains strictly larger than $\Er(R,Q)$. To the best of our knowledge, $\Etrccclb$ is the largest known threshold above which the exponent distribution for \ac{CQ} channels with arbitrary states accumulates, that is, the probability of observing a code with an exponent smaller than $\Etrccclb$ vanishes with the blocklength. Our first result also implies that, at low rates, most codes achieve an error exponent strictly larger than the random coding exponent, which is dominated by a vanishing fraction of poorly performing codes. Furthermore, we showed that, in the high-rate regime, the distribution of the exponents concentrates sharply around its mean, which implies that generating a random code with an exponent significantly different from $\Er(R,Q)$ is extremely unlikely. Another implication of this fact is that, in the same high-rate region, our bound $\Etrccclb$ is tight and coincides with the \ac{TRC} exponent. 

Our second result shows that, for any ensemble with rate $R$, a randomly selected code contains, with probability approaching $1$, a nested subcode of the same rate whose codewords all achieve the quantum expurgated exponent. This significantly strengthens Holevo’s achievability result \cite{Q_expur_general_Holevo200}, which shows that at least one such code exists. Furthermore, our second result extends Holevo's in showing that the expurgated-achieving subcode can be obtained by removing only a vanishing fraction of codewords, rather than expurgating half of them \cite{Q_expur_general_Holevo200}. Similarly as for the first result, also in the second  the gap between typical and average performance is explained by a negligible subset of poorly performing codewords.

Together, these findings demonstrate the superiority of typical (i.e.,  most) codes and codewords over ensemble averages and provide a sharper characterization of random code ensembles for \ac{CQ} channels in the asymptotic regime.
\newpage
 
\section*{Appendix}
\subsection{Proof of Lemma \ref{lemma:ex}}
We start by bounding from above the tilted error probability:
\begin{align}\label{eq:ex_chain}
 \Pecn^\frac{1}{\Grho} &= \left(\frac{1}{M_n}\sum_{m=1}^{M_n}\PecnmallMdet \right)^\frac{1}{\Grho}\\
 &\leq \frac{1}{M_n^\frac{1}{\Grho}}\sum_{m=1}^{M_n}\PecnmallMdet ^\frac{1}{\Grho}\label{eq:ex_chain_2}\\
 &\leq \frac{1}{M_n^\frac{1}{\Grho}}\sum_{m=1}^{M_n}  \sum_{m'\neq m} \left( {\rm Tr}\left\{\sqrt{\sigma^{\x_{m}}}\sqrt{\sigma^{\x_{m'}}}\right\}\right)^\frac{1}{\Grho}\label{eq:ex_chain_3}
\end{align}
where \eqref{eq:ex_chain_2} is because $(\sum_ia_i)^s\leq \sum_ia_i^s$ for $0<s\leq 1$ \cite[Ch. 5]{gallagerBook} while \eqref{eq:ex_chain_3} follows from Holevo's bound on $\PecnmallMdet^{\frac{1}{\Grho}}$ \cite{Q_expur_general_Holevo200}.
Taking the ensemble average of  \eqref{eq:ex_chain_3} and using the fact that codewords are symbol-wise i.i.d.  we have:
\begin{align}\label{eq:pe_ex}
\mathbb{E}\left[\Pecn^\frac{1}{\Grho} \right] &\leq \frac{1}{M_n^\frac{1}{\Grho}} \mathbb{E}\left[\sum_{m=1}^{M_n}  \sum_{m'\neq m} \left( {\rm Tr}\left\{\sqrt{\sigma^{\x_{m}}}\sqrt{\sigma^{\x_{m'}}}\right\}\right)^\frac{1}{\Grho}\right]\\
=& \frac{1}{M_n^\frac{1}{\Grho}}M_n(M_n-1) \mathbb{E}\left[\left( {\rm Tr}\left\{\sqrt{\sigma^{\x}}\sqrt{\sigma^{\x'}}\right\}\right)^\frac{1}{\Grho}\right]\\
=& \frac{1}{M_n^\frac{1}{\Grho}}M_n(M_n-1)\sum_{\x\in\mathcal{X}^n}\sum_{\x'\in\mathcal{X}^n} Q^n(\x)Q^n(\x')\left( {\rm Tr}\left\{\sqrt{\sigma^{\x}}\sqrt{\sigma^{\x'}}\right\}\right)^\frac{1}{\Grho}\\\label{eq:pe_ex_last}
\leq& \frac{1}{M_n^\frac{1}{\Grho}}M_n(M_n-1)\left(\sum_{x\in\mathcal{X}}\sum_{x'\in\mathcal{X}} Q(x)Q(x')\left( {\rm Tr}\left\{\sqrt{\sigma^{x}}\sqrt{\sigma^{x'}}\right\}\right)^\frac{1}{\Grho}\right)^n
\end{align}
where \eqref{eq:pe_ex_last} follows from the i.i.d. assumption \cite{Q_expur_general_Holevo200}.\unskip\nobreak\hfill$\square$

\subsection{Proof of Lemma \ref{lemma:R0}}
We proceed to Taylor-expand $\Exqn$ around $1/{\Grho}_n\rightarrow 0$. In this way we can evaluate how fast $\hat{\Grho}_n$ grows with respect to $\frac{\log \gamn}{n}$, which allows to understand whether or not $\delta_n(\hat{\Grho}_n)$ diverges.
Let us define $\sr=1/\Grho_n$. The Taylor expansion of $\Exqs$ is:
\begin{align}
    \Exqs = \nu_0(Q) - \sr{\nu_1(Q)} + O(\sr^{2})
\end{align}
where
\begin{align}
    \nu_0(Q) = \left.\Exqs \right|_{\sr=0} = -\EE\left[\log {\rm Tr}\left\{\sqrt{\sigma^{x}}\sqrt{\sigma^{x'}}\right\}\right]
\end{align}
and
\begin{align}
    \nu_1(Q) = -\left.\frac{d\Exqs}{d\sr} \right|_{\sr=0} = \frac{1}{2}\rm{Var}\left[\log {\rm Tr}\left\{\sqrt{\sigma^{x}}\sqrt{\sigma^{x'}}\right\}\right]
\end{align}
are constant in $n$.
Thus, taking the second term in \eqref{eq:exp_ineq_end_p1}, which we want to maximize, we have:
\begin{align}\notag
   &R - \frac{1}{\sr}2R + \Exqs + \delta_n(\sr) \\\notag
   &= R - \frac{1}{\sr}2R + \nu_0(Q) - \sr{\nu_1(Q)} + O(\sr^2) -\frac{1}{\sr}\frac{\log \gamn}{n}\\
   &= R - \frac{1}{\sr}\left(2R + {\sr}{O(\sr^2)}\right) + \nu_0(Q) - \sr{\nu_1(Q)}  -\frac{1}{\sr}\frac{\log \gamn}{n}\notag\\
   &\simeq R - \frac{1}{\sr}2R + \nu_0(Q) - \sr{\nu_1(Q)}  - \frac{1}{\sr}\frac{\log \gamn}{n}\label{eq:taylor3_1}
\end{align}
where the approximation in \eqref{eq:taylor3_1} becomes arbitrarily tight as $\sr$ approaches zero. Expression \eqref{eq:taylor3_1} is maximized by: 
\begin{align}\label{eq:topt}
    \hat{\sr} = \sqrt{\frac{2 \frac{\log M_n}{n} + \frac{\log \gamn}{n}}{\nu_1(Q) }},
\end{align}
where we used the definition $R=\frac{\log M_n}{n}$. 
Substituting $\hat{\sr}=1/\hat{\Grho}_n$ in \eqref{eq:topt}, we find the (asymptotic) maximizer of the second term in \eqref{eq:exp_ineq_end_p1}:
\begin{align}\label{eq:rn_opt}
    \hat{\Grho}_n = \sqrt{\frac{\nu_1(Q)}{2 \frac{\log M_n}{n} + \frac{\log \gamn}{n} }}.
\end{align}
Plugging \eqref{eq:rn_opt} in the expression $\delta_n(\Grho_n) + R - \Grho_n 2R + \Exqn$ in \eqref{eq:exp_ineq_end_p1}, we obtain a value which converges to the maximum as $n\rightarrow\infty$. If  $R=\frac{\log M_n}{n}\rightarrow 0$ and $\nu_1(Q)$ is finite, $\hat{\Grho}_n$ grows proportionally to $\sqrt{\frac{n}{\log\gamn + 2\log M_n}}<\sqrt{\frac{n}{\log\gamn}}$, which implies that, for $R=0$, $\delta_n(\hat{\Grho}_n)\rightarrow 0$ as $n\rightarrow\infty$.\unskip\nobreak\hfill$\square$

\flushend
\bibliographystyle{IEEEtran}
\bibliography{IEEEabrv,Cocco2026_Properties_CQ_Ensembles_Bib_Submit}

\end{document}